\documentclass[apj]{emulateapj}
\usepackage{amsmath}
\usepackage{apjfonts}
\usepackage{graphicx}

\newcommand{\Upa}{\ensuremath{\Upsilon}}

\newcommand{\Msun}{\ensuremath{M_{\odot}}}
\newcommand{\Lsun}{\ensuremath{L_{\odot}}}
\newcommand{\Om}{\ensuremath{\Omega_M}}
\newcommand{\Omh}{\ensuremath{\Omega_M h}}
\newcommand{\ergscm}{erg~s$^{-1}$~cm$^{-2}$}

\begin{document}

\shorttitle{BARYON MASS FUNCTION}
\shortauthors{VOEVODKIN \& VIKHLININ}
\slugcomment{ApJ, in press ({\normalfont astro-ph/0305549})}

\title{Constraining amplitude and slope of the mass fluctuation spectrum\\
  using cluster baryon mass function.}

\author{A. Voevodkin\altaffilmark{1}, A.
  Vikhlinin\altaffilmark{2}\altaffilmark{,1}}

\altaffiltext{1}{Space Research Institute, Profsoyuznaya 84/32, Moscow,
  Russia.\\
  voevodkin@hea.iki.rssi.ru}
\altaffiltext{2}{Harvard-Smithsonian Center for Astrophysics, 60
  Garden St., Cambridge, MA 02138}

\begin{abstract}
  We derive the baryon mass function for a complete sample of low-redshift
  clusters and argue that it is an excellent proxy for the total mass
  function if the ratio $f_b=M_b/M_{\rm tot}$ in all clusters is close to
  its universal value, $\Omega_b/\Om$. Under this assumption, the baryon
  mass function can be used to constrain the amplitude and slope of the
  density fluctuations power spectrum on cluster scales. This method does
  not use observational determinations of the total mass and thus bypasses
  major uncertainties in the traditional analyses based on the X-ray
  temperature function. However, it is sensitive to possible systematic
  variations of the baryon fraction as a function of cluster mass.  Adapting
  a weak dependence $f_b(M)$ suggested by observations and numerical
  simulations by Bialek et al., we derive $\sigma_8=0.72\pm0.04$ and the
  shape parameter $\Omh=0.13\pm0.07$, in good agreement with a number of
  independent methods. We discuss the sensitivity of these values to other
  cosmological parameters and to different assumptions about variations in
  $f_b$.
\end{abstract}

\keywords{cosmological parameters --- galaxies: clusters: general ---
surveys --- X-rays: galaxies}

\section{Introduction}

The number density of galaxy clusters, as a function of mass, is an
excellent indicator for the normalization and slope of the power spectrum of
the density fluctuations in the present-day Universe. The power spectrum
normalization is traditionally expressed in terms of $\sigma_8$, the
\emph{rms} amplitude of the density fluctuations on the $8\,h^{-1}$~Mpc
scale.  In the CDM cosmology, the shape of the present-day power spectrum is
controlled almost completely by the product $\Omh$ (Bond \& Efstathiou
1984). Therefore, the cluster mass function is a sensitive probe of these
cosmological parameters.

Considerable progress in the theoretical models of structure formation in
the past decade has provided a reliable machinery to compute the shape of
the power spectrum for any given set of cosmological parameters (Seljak \&
Zaldarriaga 1996, Eisenstein \& Hu 1998), and to predict the mass function
of collapsed objects (clusters) given the power spectrum (Jenkins et al.\
2001). Most of the problems with using the cluster mass function as a
cosmological probe are on the observational side.

Theory operates with the virial masses of clusters. However, the total
cluster mass within the virial radius, as opposed to the mass within a small
metric radius, is notoriously difficult to measure. The main techniques such
as weak lensing, the X-ray method assuming hydrostatic equilibrium of the
intracluster gas, and velocity dispersion of cluster galaxies, lead to
significant statistical and systematic uncertainties at large radii, and
often produce contradictory results (see e.g., Markevitch \& Vikhlinin 1997
and Fischer \& Tyson 1997 for representative uncertainty estimates for the
X-ray hydrostatic and weak lensing methods, respectively). It is probably
fair to say that total masses at the virial radius cannot be determined to
better than 30--40\% accuracy at present.

Because of the difficulties with the virial mass measurement, a direct
derivation of the mass function for a large, complete sample of clusters is
not feasible at present. Therefore, the cluster mass is usually related to a
more easily measured quantity. Consider one of the most frequently used
proxies for the mass function, the cluster temperature function. This relies
on the theoretically expected tight correlation $M\propto T^{3/2}$. The
theoretically expected slope of the $M-T$ correlation is reproduced both in
numerical simulations (Evrard, Metzler \& Navarro 1996) and observationally,
within a given approach to the total mass measurement (Horner, Mushotzky, \&
Scharf 1999; Nevalainen, Markevitch \& Forman 2000; Finoguenov, Reiprich \&
B\"ohringer 2001). However, the normalizations of the $M-T$ relation derived
by different authors span a wide range (see Evrard et al.\ 2002 for a
review).  Ultimately, the question of the normalization of the $M-T$
relation boils down to the possibility of total mass measurements with small
systematic uncertainties, which is at the limits of current observational
techniques.

The expected cluster mass functions are very steep and therefore any change
in the mass scale leads to large variations of the derived cosmological
parameters. In a recent review, Rosati et al.\ (2002) estimated that 30\%
uncertainties in the normalization of the $M-T$ relation leads to a 20\%
uncertainties in $\sigma_8$ and 30\% uncertainties in \Om{} derived from cluster
evolution. Until the $M-T$ relation is reliably normalized, precise
measurements of the cosmological power spectrum from the cluster temperature
function are impossible. Therefore, it is desirable to explore new proxies for
the cluster mass function that would not require absolute measurements of
the total mass. In this Paper, we propose that such a proxy can be based on
the easily measured cluster baryon masses.

Indeed, it is expected that the fraction of baryons in the virial mass of
clusters equals the average value in the Universe --- $f_b \equiv
M_b/M_{\text{tot}} = \Omega_b/\Om$ (White et al.\ 1993). Therefore, the
baryon fraction is the same for clusters of any mass at any redshift at
least to a first approximation. This is one of the best-proved
cluster-related facts, that is justified by basic theory, confirmed by
numerical simulations (e.g., Frenk et al.\ 1998), and supported
observationally (Mohr, Mathiesen \& Evrard 1999, Allen et al. 2002). Given
the universality of the baryon fraction, there is a trivial relation between
the cumulative total and baryon mass functions:
\begin{equation}
F_b(M_b) = F_{\text{tot}}(\Om/\Omega_b M_b).
\label{eq:Fb-Ft}
\end{equation}
The average baryon density in this equation is fixed by either the Big Bang
Nucleosynthesis theory or by the cosmic microwave background fluctuations,
$\Omega_b h^2 = 0.0224\pm0.0009$ (Spergel et al.\ 2003). The baryon mass is
a convenient proxy for the total mass because 1) it is easily derived from
the X-ray imaging observations (\S\,\ref{sec:M_b:meas}), and 2) the scaling
between $M_b$ and $M_{\text{tot}}$ is expressed in terms of the model
parameters ($\Om$) and therefore does not rely on the absolute total mass
measurements. In this Paper, we apply this method to \emph{ROSAT}
observations of a complete flux-limited sample of 52 low-redshift clusters
selected from the All-Sky Survey data.

The Paper is organized as follows. In \S\,\ref{sec:cluster-sample}, we
describe selection of our cluster sample. The measurements of the baryon
masses are presented in \S\,\ref{sec:M_b:meas} and the baryon mass function
is derived in \S\,\ref{sec:mfun}. The theory involved in modeling the
baryon mass function is reviewed in \S\,\ref{sec:theory}. In
\S\,\ref{sec:results}, we derive constraints on $\sigma_8$ and on the shape
parameter, $\Gamma=\Omh$, and discuss the dependence of these results
on our basic assumption of the universality of the baryon fraction.

Numerical values of the cluster parameters are quoted for the
$H_0=71$~km~s$^{-1}$~Mpc$^{-1}$, $\Omega_M=0.3$ and $\Lambda=0.7$ cosmology.

\section{Cluster Sample}
\label{sec:cluster-sample}

\begin{table*}
  \def\tnote#1{\ensuremath{^{\text{#1}}}}
  \def\d{\phantom{1}}
    \centering
    \caption{Cluster sample}\label{tab:clust.par}
    \medskip\def\arraystretch{1.15}
    \footnotesize
    \begin{tabular}{lcccccccc}
      \hline
      \hline
      \multicolumn{1}{c}{Object} & $z$ & $T$\tnote{\it a}& $f_x$\tnote{\it
        b} & $L_x$ & $M_{g,324}$ & $M_{b,324}$ & $r_{g,324}$ & $L_{\text{opt}}$\tnote{\it c}\\
       & & (keV) & & $10^{44}$~erg~s$^{-1}$ & $10^{14}\Msun$ & $10^{14}\Msun$
       & (Mpc) & $10^{13}\Lsun$\\
      \hline
2A0335  & 0.0349 & 3.6&  6.8&  1.8& $0.59\pm0.07$ & $0.78\pm0.10$ & 1.85 & \nodata \\
A85     & 0.0556 & 6.9&  4.3&  3.0& $1.11\pm0.13$ & $1.35\pm0.16$ & 2.24 &$0.41\pm0.06$\\
A119    & 0.0442 & 5.6&  2.4&  1.0& $1.00\pm0.10$ & $1.27\pm0.13$ & 2.19 & \nodata \\
A262    & 0.0155 & 2.2&  4.6&  0.2& $0.20\pm0.04$ & $0.31\pm0.06$ & 1.31 &$0.09\pm0.03$\\
A399    & 0.0715 & 6.8&  1.7&  2.0& $1.21\pm0.31$ & $1.62\pm0.38$ & 2.27 & \nodata \\
A400    & 0.0238 & 2.3&  1.6&  0.2& $0.25\pm0.02$ & $0.38\pm0.04$ & 1.41 &$0.11\pm0.03$\\
A401    & 0.0739 & 8.0&  3.1&  3.9& $2.34\pm0.26$ & $2.82\pm0.31$ & 2.82 &$0.47\pm0.18$\\
A478    & 0.0882 & 6.9&  4.5&  8.2& $2.16\pm0.24$ & $2.59\pm0.29$ & 2.71 &$0.36\pm0.09$\\
A496    & 0.0328 & 4.7&  5.4&  1.3& $0.61\pm0.08$ & $0.80\pm0.11$ & 1.87 & \nodata \\
A576    & 0.0389 & 4.0&  1.6&  0.5& \nodata       & $0.46\pm0.14$\tnote{*} &\nodata& \nodata \\
A754    & 0.0542 & 9.8&  4.3&  2.9& $1.58\pm0.16$ & $1.90\pm0.20$ & 2.52 & \nodata \\
A1060   & 0.0137 & 3.2&  6.1&  0.2& $0.20\pm0.04$ & $0.29\pm0.06$ & 1.31 & \nodata \\
A1367   & 0.0214 & 3.7&  4.7&  0.5& $0.41\pm0.07$ & $0.54\pm0.09$ & 1.66 &$0.14\pm0.04$\\
A1644   & 0.0474 & 4.7&  2.2&  1.1& $0.92\pm0.21$ & $1.19\pm0.25$ & 2.11 & \nodata \\
A1651   & 0.0860 & 6.2&  1.5&  2.6& $1.14\pm0.17$ & $1.42\pm0.21$ & 2.19 & \nodata \\
A1656   & 0.0231 & 8.1& 18.2&  2.1& $1.54\pm0.18$ & $1.91\pm0.23$ & 2.57 &$0.51\pm0.11$\\
A1736   & 0.0461 & 3.6&  1.7&  0.8& \nodata       & $0.66\pm0.20$\tnote{*} &\nodata& \nodata \\
A1795   & 0.0622 & 6.0&  3.9&  3.4& $0.93\pm0.08$ & $1.15\pm0.11$ & 2.10 &$0.25\pm0.04$\\
A2029   & 0.0766 & 9.1&  4.3&  5.7& $1.86\pm0.24$ & $2.30\pm0.29$ & 2.60 &$0.58\pm0.12$\\
A2052   & 0.0353 & 3.4&  3.0&  0.8& $0.31\pm0.05$ & $0.42\pm0.07$ & 1.49 &$0.13\pm0.03$\\
A2063   & 0.0355 & 3.7&  2.5&  0.7& $0.44\pm0.07$ & $0.60\pm0.09$ & 1.68 &$0.17\pm0.03$\\
A2065   & 0.0726 & 5.4&  1.4&  1.7& \nodata       & $1.13\pm0.34$\tnote{*} &\nodata& \nodata \\
A2142   & 0.0894 & 9.3&  3.9&  7.3& $3.13\pm0.23$ & $3.73\pm0.28$ & 3.06 &$0.56\pm0.09$\\
A2147   & 0.0353 & 4.4&  3.1&  0.9& $1.01\pm0.18$ & $1.33\pm0.22$ & 2.21 & \nodata \\
A2163   & 0.2010 &11.5&  1.4& 14.6& $5.58\pm0.74$ & $6.41\pm0.85$ & 3.37 & \nodata \\
A2199   & 0.0299 & 4.8&  6.5&  1.3& $0.57\pm0.04$ & $0.75\pm0.06$ & 1.84 &$0.24\pm0.04$\\
A2204   & 0.1523 & 7.1&  1.6&  9.4& $2.07\pm0.43$ & $2.51\pm0.49$ & 2.52 & \nodata \\
A2256   & 0.0581 & 7.3&  3.7&  2.8& $1.48\pm0.12$ & $1.80\pm0.15$ & 2.45 &$0.54\pm0.06$\\
A2589   & 0.0416 & 3.7&  1.6&  0.6& $0.29\pm0.04$ & $0.42\pm0.06$ & 1.46 & \nodata \\
A2634   & 0.0309 & 3.3&  1.5&  0.3& $0.31\pm0.04$ & $0.42\pm0.05$ & 1.50 &$0.19\pm0.03$\\
A2657   & 0.0400 & 3.7&  1.7&  0.6& $0.40\pm0.05$ & $0.54\pm0.06$ & 1.62 & \nodata \\
A3112   & 0.0750 & 5.3&  1.9&  2.5& $0.81\pm0.13$ & $1.03\pm0.16$ & 1.98 & \nodata \\
A3158   & 0.0591 & 5.8&  2.3&  1.8& $0.90\pm0.15$ & $1.17\pm0.19$ & 2.09 & \nodata \\
A3266   & 0.0589 & 8.0&  3.5&  2.8& $1.71\pm0.17$ & $2.05\pm0.20$ & 2.57 & \nodata \\
A3376   & 0.0455 & 4.0&  1.4&  0.6& $0.41\pm0.06$ & $0.54\pm0.07$ & 1.63 & \nodata \\
A3391   & 0.0514 & 5.7&  1.5&  0.9& $0.93\pm0.14$ & $1.24\pm0.18$ & 2.12 & \nodata \\
A3395   & 0.0506 & 5.0&  1.9&  1.1& $1.24\pm0.20$ & $1.59\pm0.25$ & 2.33 & \nodata \\
A3526   & 0.0112 & 3.4& 15.7&  0.4& $0.31\pm0.05$ & $0.44\pm0.07$ & 1.51 & \nodata \\
A3558   & 0.0480 & 5.5&  4.0&  2.1& $1.34\pm0.12$ & $1.66\pm0.16$ & 2.39 & \nodata \\
A3562   & 0.0490 & 5.2&  1.8&  1.0& $0.92\pm0.09$ & $1.17\pm0.13$ & 2.12 & \nodata \\
A3571   & 0.0391 & 7.2&  7.5&  2.5& $1.21\pm0.18$ & $1.50\pm0.22$ & 2.33 & \nodata \\
A3581   & 0.0214 & 1.8&  2.0&  0.2& \nodata       & $0.23\pm0.07$\tnote{*} &\nodata& \nodata \\
A3667   & 0.0556 & 7.0&  4.6&  3.2& $2.50\pm0.24$ & $2.99\pm0.29$ & 2.93 & \nodata \\
A4038   & 0.0292 & 3.3&  3.6&  0.7& $0.31\pm0.05$ & $0.44\pm0.07$ & 1.50 & \nodata \\
A4059   & 0.0460 & 4.4&  2.0&  0.9& $0.45\pm0.07$ & $0.62\pm0.09$ & 1.67 & \nodata \\
EXO 0422& 0.0390 & 2.9&  1.9&  0.6& \nodata       & $0.53\pm0.16$\tnote{*} &\nodata& \nodata \\
Hydra A & 0.0522 & 3.6&  2.9&  1.8& $0.60\pm0.07$ & $0.77\pm0.09$ & 1.83 & \nodata \\
MKW 3s  & 0.0453 & 3.7&  2.1&  1.0& $0.40\pm0.06$ & $0.53\pm0.08$ & 1.60 &\nodata \\
MKW 4   & 0.0210 & 1.7&  1.4&  0.1& $0.14\pm0.01$ & $0.21\pm0.03$ & 1.15 &$0.05\pm0.02$\\
NGC1550 & 0.0120 & 1.1&  2.5&  0.1& $0.05\pm0.01$ & $0.08\pm0.02$ & 0.84 & \nodata \\
NGC507  & 0.0155 & 1.3&  1.5&  0.1& $0.08\pm0.01$ & $0.13\pm0.02$ & 0.97 & \nodata \\
S 1101  & 0.0580 & 2.6&  1.5&  1.1& $0.29\pm0.04$ & $0.39\pm0.06$ & 1.43 & \nodata \\
\hline
\end{tabular}

\bigskip

\begin{minipage}{0.8\linewidth}

$^a$ Cluster temperatures adapted from Markevitch et al. (1998,
  1999), White et al. (2000), Fukazawa et al. (1998), David et al. (1993),
  Arnaud et al. (2001), Blanton et al. (2001), Horner et al. (1999), Allen
  \& Fabian (1998), Finoguenov et al. (2001), Kaastra et al. (2001), Sun et
  al. (2003)

$^b$ X-ray flux in 0.5--2.0 keV, in $10^{-11}$\ergscm\ units.

$^c$ Optical luminosity inside $r_{g,324}$ in the V band, adapted
  from Arnaud et al. (1992), Hradecky et al.  (2000), Beers et al. (1984).

$^*$ --- baryon mass determined from $M_{b,324}-L_x$ correlation.
\end{minipage}

\end{table*}

Our cluster sample was selected from the HIFLUGCS catalog (Reiprich \&
B\"ohringer 2002) of clusters selected from the \emph{ROSAT} All-Sky Survey
(RASS) data. This catalog includes 63 clusters above a flux limit of
$2\times10^{-11}$~erg~s$^{-1}$~cm$^{-2}$ in the 0.1--2.4~keV band, detected
over 2/3 of the sky.

We have applied further objective selections to facilitate a more reliable
measurement of the baryon mass function. First, we increased the flux limit
to $1.4\times10^{-11}$~\ergscm~in the 0.5--2~keV band (or
$2.3\times10^{-11}$\ergscm{} in the 0.1--2.4~keV band). Second, we considered
only clusters at $z>0.01$ because at lower redshifts, the virial radius of
most objects does not fit inside the \emph{ROSAT} field of view. In total,
52 clusters from the HIFLUGCS catalog satisfied our selection criteria
(Table~\ref{tab:clust.par}).

Most of the selected objects were observed by \emph{ROSAT} in one or more
pointed observations, so high-quality X-ray data are available.  Several
more clusters are so bright that their baryon mass can be measured from the
shallower RASS data. For only 5 clusters, A576, A1736, A2065, A3581, and
EXO422, the existing X-ray data do not permit a direct baryon mass
measurement at the virial radius, so we used the $M_b-L_x$ correlation to
estimate their masses.

\section{Baryon mass measurement}
\label{sec:M_b:meas}

\subsection{\emph{ROSAT} Data Reduction and Gas Mass Measurements}
\label{sec:mgas}

The data were prepared in a standard manner. For pointed observations, we
used S.~Snowden's software (Snowden et al.\ 1994) to produce flat-fielded
images with all non-sky background components removed. For the RASS data, we
used the archive-supplied exposure maps. Experiments with different energy
bands show that the 0.7--2~keV band has the optimal ratio of cluster to
background surface brightness. Our further analysis is performed in this
band.

The measurement of the gas mass profile from the X-ray imaging data was
performed almost identically to the approach described in Vikhlinin, Forman
\& Jones (1999). We first extracted the azimuthally-averaged surface
brightness profile centered on the apparent centroid of the cluster. The
background level was estimated by fitting the profile at large radii, $0.3
r_v < r < 1.5 r_v$ where $r_v$ is the virial radius estimated from the X-ray
temperature\footnote{We used the relation provided by Evrard, Metzler \&
  Navarro (1996), $r_v = 1.95\,
  h^{-1}\,\mbox{Mpc}\,(T/10\,\mbox{keV})^{1/2}$}, with the power law plus
constant model. The power law here represents the cluster brightness and the
constant corresponds to the background. The background intensity was
subtracted from the profile assuming that it is constant across the image,

Assuming that the clusters are spherically symmetric, the gas density
profiles were derived from the surface brightness profiles using direct
deprojection (Fabian et. al. 1981). The deprojection method yields the
radial profile of the X-ray volume emissivity which is easily converted to
the gas density as $n_e=K\,\varepsilon^{1/2}$, $\rho_g = 1.15 n_e m_p$,
where the numerical value for $K$ can be found for any plasma temperature
using the MEKAL code (Mewe et al. 1985), and the numerical value in the
second relation corresponds to the standard mix of fully ionized hydrogen
and helium.

For three clusters, A1060, A1644, and NGC1550, the \emph{ROSAT} surface
brightness profile were too noisy for direct deprojection. We resorted to
the usual $\beta$-model analysis in these three cases.

The intracluster gas which we directly observe in X-rays is the dominant
baryonic component of clusters. We will use the cluster masses corresponding
to the density contrast $\delta=324$ relative to the mean density of the
Universe at the redshift of observation (see \S\,\ref{sec:theory}
below). Therefore, we derive the gas masses from the following equation: $3
M_g(<\!r)/(4\pi r^3\,\langle\rho_b\rangle)=324$, where $M_g(<\!r)$ is the
gas mass profile as derived from the X-ray data.  We adopt the average
baryon density as given by the \emph{WMAP} observations of the cosmic microwave
background fluctuations, $\langle\rho_b\rangle=
6.23\,(1+z)^3$\,\Msun~kpc$^{-3}$ (or $\Omega_b\,h^2=0.0224$, Spergel et al.\
2003) which is also close to the Big Bang Nucleosynthesis value of
$\Omega_b\,h^2=0.020$ (Burles, Nollett \& Turner 2001). The observed X-ray
surface brightness profiles permit a direct measurement of the gas mass at
the radius corresponding to $\delta=324$.

The measured gas masses $M_{g,324}$ are listed in
Table~\ref{tab:clust.par}. It is useful to provide simple relations which
allow one to scale our results to other values of the mean overdensity,
Hubble constant etc. Empirically, we find that on average, the radial
dependence of mass is $M(\theta)\propto \theta$, which corresponds to the
following scaling as a function of overdensity
\begin{equation}\label{eq:m:delta}
  M_\delta \propto \delta^{-0.5}.
\end{equation}
Within a fixed aperture, the X-ray derived baryon mass scales as
$h^{-5/2}$. There is an additional scaling because $M_\delta$ is measured at
a fixed overdensity $\delta$ rather than within a fixed angular radius. The
angular radius corresponding to the overdensity $\delta$ is found from the
equation $M(\theta_\delta)/(\theta_\delta^3 h^{-3})=\mathrm{const}$. Given
the empirical average mass profile, $M(\theta)\propto \theta$, and the
$h$-scaling of $M(\theta)$, $\theta_\delta$ should scale as $h^{0.25}$, and
therefore
\begin{equation}\label{eq:m:h}
  M_\delta \propto h^{-2.25}.
\end{equation}
The only effect of changing the average baryon density in the Universe is to
change the apparent mean overdensity, and so one can use
eq.~(\ref{eq:m:delta}) to convert our masses to $\delta=324$ for a different
value of $\Omega_b\,h^2$:
\begin{equation}\label{eq:m:rhob}
  M_\delta \propto (\Omega_b h^2)^{-0.5}.
\end{equation}

\begin{figure}
  \vspace*{-5mm}
  \centerline{
    \includegraphics[width=0.99\linewidth]{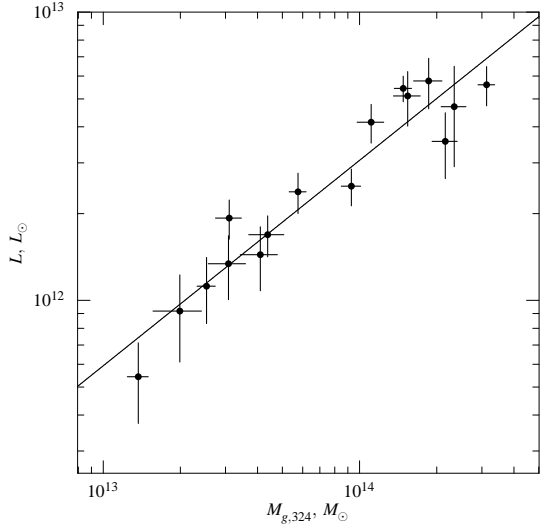} } \vspace*{-7mm}
  \caption{Correlation of the optical luminosity of the cluster in the V
    band and the gas mass ($M_{g,324}$). The original optical measurements
    were extrapolated to $r_{g,324}$ as explained in the text. The solid
    line shows the best-fit power law (eq.~\ref{eq:lopt-mgas})}
  \label{fig:Lopt-Mgas}
\end{figure}

\subsection{Stellar Mass in Clusters}
\label{sec:mstars}

The stellar material in galaxies contributes a small but non-negligible
fraction of the cluster baryon mass. Not all clusters in our sample have
high-quality optical observations. Therefore, we first use the published
data for a subsample of our clusters to establish a correlation of the
optical luminosity with the gas mass and then use this correlation to
estimate the stellar contribution in all clusters.

The optical luminosities for 16 clusters were compiled from the papers by
Arnaud et al.\ (1992), Hradecky et al.\ (2000), and Beers et al. (1984). The
optical luminosities in these papers are quoted at different metric radii in
the V band, and so we converted them to the gas overdensity radius $r_{324}$
assuming that the light follows the King profile with $r_{c} = 180$~kpc
(Bahcall 1975). The resulting luminosities within $r_{324}$ show a very
tight correlation with the gas mass (Fig.~\ref{fig:Lopt-Mgas}) which is
fitted by the following relation
\begin{equation}\label{eq:lopt-mgas}
L_{\text{opt}} = 319\Lsun\,\bigg[\frac{M_{g,324}}{M_\odot} \bigg]^{0.71}
\end{equation}

We further assume the same mass-to-light ratio for the stars,
$M_{*}/L_{\text{opt}} = 6\Msun/\Lsun$ in the V band, which follows from the
stellar evolution models for the average population of elliptical and spiral
galaxies in clusters (Arnaud et. al. 1992). Under these assumptions,
equation~[\ref{eq:lopt-mgas}] can be used to estimate the total baryon
(i.e.\ gas plus galactic stars) mass of a cluster from the X-ray measured
gas mass:
\begin{equation}
\frac{M_{b,324}}{M_{g,324}} = 1.100 + 0.045\,\biggl[\frac{M_{g,324}}{10^{15}\Msun}\biggl]^{-0.5}.
\end{equation}
Therefore, galactic stars contribute about 15\% of the total baryon mass in
massive clusters, and this fraction slightly increases in the less massive
systems. This estimate of the stellar contribution is consistent with the
considerations presented in Fukugita, Hogan \& Peebles (1998). The trend in
the stellar-to-gas ratio is similar to that derived by Lin, Mohr \& Stanford
(2003) using the $K$-band photometry from the 2MASS survey, although the
overall normalization of $M_*/M_g$ is smaller in that paper.

We ignore any contribution from intergalactic stars to the cluster baryon
mass. At present this component cannot be reliably estimated. However, all
available studies (e.g., Feldmeier et al.\ 2002) suggest that the
intergalactic stars contribute only a small fraction to the total stellar
mass of the cluster, and therefore only a few percent at most to the total
baryon mass.

\subsection{Baryon Mass Error Budget}
\label{sec:errorbudget}

The main contribution to the mass measurement uncertainty is the Poisson
noise in the X-ray surface brightness profiles. This noise component is
straightforward to compute in each case. Typically, the statistical
uncertainty in the baryon mass is about 20\%. All other uncertainties are
systematic and we can only estimate their contribution.

We measured the baryon masses under the assumption that the clusters are
spherical. Any deviations from the spherical symmetry result in a scatter in
our mass determination. We estimate this scatter to be 7\% based on the
azimuthal scatter of the gas mass profiles (Vikhlinin et al.\ 1999) and from
the results of numerical simulations by Mathiesen, Evrard \& Mohr (1999).

The stellar mass was estimated using a number of assumptions. First, we
assumed a standard mass-to-light ratio for the galactic stars. Second, the
total light was estimated from the extrapolation of the King profile to
large radii. Third, for most clusters, the optical luminosity was estimated
from the $L_{\text{opt}}-M_{\text{g}}$ relation which has some scatter
(Fig.~\ref{fig:Lopt-Mgas}). We conservatively estimate that the stellar mass
is measured with a 30\% uncertainty. The stellar mass is only 15--20\% of
the gas mass, therefore the contribution to the uncertainty in the total
baryon mass is small (5--7\%).

The final uncertainties on the baryon mass were obtained by adding the above
three components in quadrature. The dominant contribution is still the
statistical noise in the X-ray data.

\begin{figure}
  \vspace*{-5mm} \centerline{
    \includegraphics[width=0.99\linewidth]{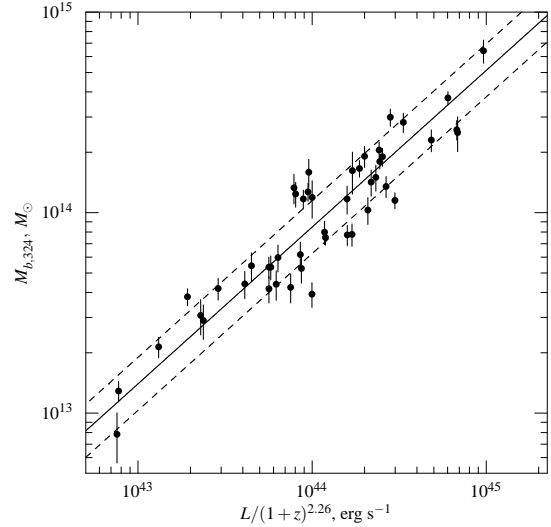} } \vspace*{-7mm}
  \caption{Cluster baryon mass as a function of the total X-ray luminosity
    in the 0.5--2 keV band. The solid line shows the best fit power law and
    dashed lines indicate the 1$\sigma$ scatter around the mean relation.}
  \label{fig:Mbar-Lx}
\end{figure}

\subsection{Correlation of the X-ray Luminosity and the Baryon Mass}
\label{sec:M-L}

The key relation for estimating the mass functions is the volume of the
survey as a function of mass. For a flux-limited sample like ours, the
volume is determined by the scaling relation between the mass and the X-ray
luminosity. Such a relation is also useful for estimating the masses in
those cases where direct mass measurements are impossible.

Our sample was selected using the total (i.e.\ without excluding either
substructures or cooling flow regions) X-ray fluxes in the 0.5--2 keV band.
Therefore, the $M_b-L_x$ correlation should be with the total luminosity in
this energy band. The corresponding measurements are listed in Table
\ref{tab:clust.par} and plotted in Fig.~\ref{fig:Mbar-Lx}. The average
$M_{b,324}-L_x$ correlation is well fitted by the power law\footnote{ The
$M_{b}-L_{x}$ relation evolves strongly to $z\sim0.5$ (Vikhlinin et al.\
2002). To correct for this effect, we divided the luminosities by the
observed rate of the evolution $(1+z)^{2.26}$ even though this correction is
small in our redshift range.}
\begin{equation}\label{eq:Mb-Lx}
M_{b,324} = 8.5\times10^{13}\Msun \times
(L_x/10^{44}\,\mathrm{erg~s^{-1}})^{0.78\pm0.04} 
\end{equation}
where the best-fit parameters were found using the bisector method of
Akritas \& Bershady (1996) and the uncertainty in the slope was estimated
using bootstrap resampling.

\begin{figure}
  \vspace*{-5mm}
  \centerline{
    \includegraphics[width=0.99\linewidth]{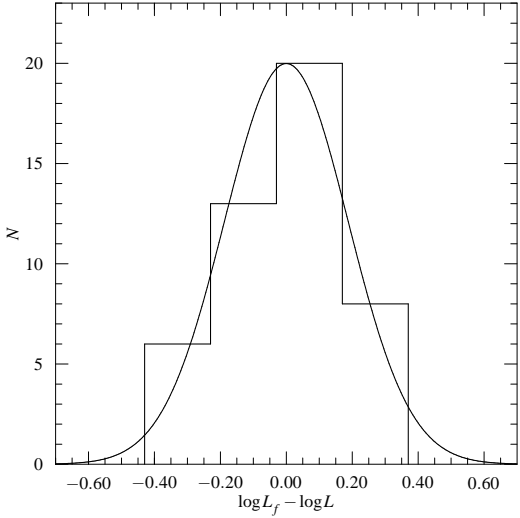} } \vspace*{-7mm}
  \caption{Deviations of the measured luminosity from the mean $M_b-L_x$
    relation. Solid line shows the Gaussian fit.}
  \label{fig:gist}
\end{figure}

There is a significant scatter in the $M_{b,324}-L_x$ correlation which can
be partly explained by the contribution of cooling flow regions and
substructures to the total cluster flux. As Figure~\ref{fig:gist} shows, the
scatter around the mean relation is well approximated by the log-normal
distribution with an \emph{rms} scatter of $\sigma_{\lg M}=0.19$ in mass.
Part of this scatter is due to the mass measurement errors. Subtracting this
contribution in quadrature, we find that the intrinsic log scatter in the
$M_{b,324}-L_x$ correlation is $\sigma_{\lg M} = 0.17$ in mass for a given
luminosity or $\sigma_{\lg L}=0.15$ in luminosity for a given mass.

\section{Observed Baryon Mass Function}\label{subsec:bfm}
\label{sec:mfun}

The volume of a flux-limited survey for the cluster of luminosity $L$ is
obtained by integration of the cosmological dependence of comoving volume,
$dV/dz$, over the redshift interval $z_{\text{min}}-z_{\text{max}}$, where
$z_{\text{min}}$ is the lower redshift cut (0.01 in our case), and
$z_{\text{max}}$ is the maximum redshift for the cluster to exceed the flux
limit, $f_{\text{lim}}$:
\begin{equation}
 f_{\text{lim}} (0.5-2) =
 \frac{L\left(0.5(1+z_{\text{max}})-2(1+z_{\text{max}})\right)}{4\pi\,
   d_L^2(z_{\text{max}})},
\end{equation}
where $d_L(z_{\text{max}})$ is the luminosity distance as a function of
redshift. The solution of this equation with respect to $z_{\text{max}}$
leads to the survey volume as a function of cluster luminosity, $V(L)$.

\begin{figure}[t]
  \vspace*{-5mm} \centerline{
    \includegraphics[width=0.99\linewidth]{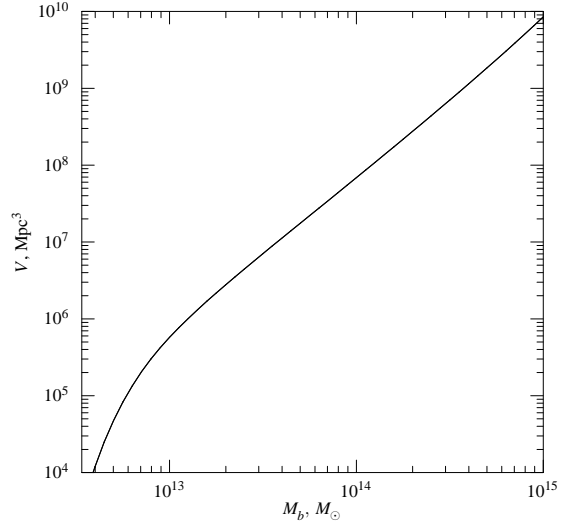} } \vspace*{-7mm}
  \caption{Sample volume as a function of the cluster baryon mass.}
  \label{fig:volume}
\end{figure}

Given the probability distribution $p(L|M)$ of the cluster luminosity for
the given mass, the survey volume as a function of mass is computed as $V(M)
= \int p(L|M) V(L)\,dL$. Taking into account the average relation between
$M_b$ and $L$ (eq.~\ref{eq:Mb-Lx}) and the log-normal scatter around this
relation, we obtain:
\begin{equation}\label{eq:volume}
  \begin{split}
    V(M) = &\frac{1}{(2 \pi)^{1/2}\sigma_{\lg M}}\; \times \\
    & \int \exp\biggl[ -\frac{(\lg L_f(M) - \lg L)^2}{2 \sigma_{\lg M}^2} \biggl]  V(L)\,d\lg L,
  \end{split}
\end{equation}
where $L_f(M)$ is given by the inverse of eq.~(\ref{eq:Mb-Lx}). The sample
volume as a function of the cluster baryon mass is shown in
Fig.~\ref{fig:volume}. 

\begin{figure}[b]
  \vspace*{-5mm} \centerline{
  \includegraphics[width=0.99\linewidth]{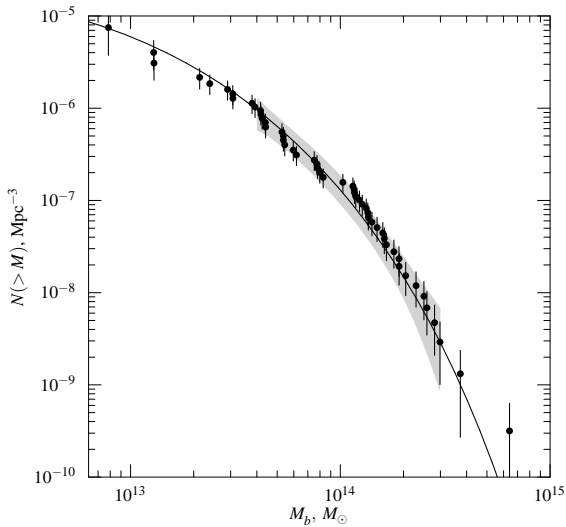} } \vspace*{-7mm}
  \caption{Local baryon mass function. The gray band shows a 68\%
    uncertainty interval derived from the Maximum Likelihood analysis
    including the mass measurement errors. The solid line corresponds to the
    prediction of our best-fit cosmological model
    (Fig.~\ref{fig:1-2sig.contour}).}
  \label{fig:cumfunc}
\end{figure}

We will use the cumulative representation of the mass function which is
estimated from observations as
\begin{equation}
  f(M) = N(>M) = \sum_{M_i\ge M}\frac{1}{V(M_i)},
\end{equation}
where $V(M_i)$ is given by eq.~[\ref{eq:volume}]. The resulting baryon mass
function is shown in Fig.~\ref{fig:cumfunc}. 

There are two contributions to the uncertainty in the $f(M)$ measurement ---
the Poisson noise which is trivially estimated as
\begin{equation}
  \left(\Delta_P f(M)\right)^2 = \sum_{M_i\ge M}\frac{1}{V^2(M_i)},
\end{equation}
and the component related to the mass measurement uncertainties. To take
both components into account, we performed the Maximum-Likelihood fits of
the measured $f(M)$ with a two-parameter Schechter function $df/dM = A
(M/M_0)^\alpha \exp(-M/M_0)$ (Schechter 1976). The likelihood function was
constructed to include the scatter due to the mass measurement errors (for
details, see Voevodkin, Vikhlinin \& Pavlinsky 2002 and Vikhlinin et al.\
2003). The range of statistically acceptable parameters $A$, $\alpha$, and
$M_0$ defines a band of the cumulative mass functions consistent with the
data at the given confidence level. The resulting 68\% uncertainties in the
$f(M)$ measurement are shown as a grey band in Fig.~\ref{fig:cumfunc}. For
reasons explained below (\S~\ref{sec:results}), we use only clusters with
baryon mass $M_b > 4\times10^{13}\Msun$ for the theoretical
modeling. Therefore, the Schechter function fits were performed only in this
mass range.

In addition to the statistical uncertainty in the mass function there is an
additional component because of the cosmic variance in a survey of a finite
volume. Hu \& Kravtsov (2003) calibrate the ratio of the cosmic variance and
Poisson noise as a function of cluster mass. Using their results, we find
that the cosmic variance is negligible in our mass range for
$\sigma_8\simeq0.7-0.8$ indicated by our data (see below).

\section{Review of Theory}
\label{sec:theory}

Recent numerical simulations of structure formation in the Universe provide
accurate analytic expressions for the cluster total mass function (Jenkins
et al.\ 2001). Jenkins et al.\ have demonstrated that the mass functions in
the CDM cosmology has a universal form when expressed as a function of the
linear \emph{rms} density fluctuations on the mass scale $M$, $\sigma(M)$:
\begin{equation}\label{eq:mfun:jenkins}
\frac{M}{\langle\rho\rangle} \frac{d N(>\!M)}{d \log \sigma(M)} = A \exp\,(-|\log \sigma^{-1}+b\,|^c)
\end{equation}
All dependencies of the mass function on cosmological parameters enter
through $\sigma(M)$. The function $\sigma(M)$ is equivalent to the power
spectrum of linear density perturbations (Peebles 1981). We will assume that
the power spectrum, $P(k)$, is the product of the primordial inflationary
spectrum $k^n$ and the CDM transfer function: $P(k)\sim k^n T^2(k)$. The
transfer function depends on the parameters \Om, $h$, and $\Omega_b$. We
used analytic approximations to $T(k)$ provided by Eisenstein \& Hu
(1998). Therefore, the theoretical model for the mass function is defined by
parameters $\Om$, $\Omega_b$, $h$, $n$, and also the normalization of the
power spectrum, $\sigma_8$.

The numerical values of coefficients $A$, $b$, and $c$ in
(\ref{eq:mfun:jenkins}) depend on the definition of the cluster mass.  The
most convenient definition, from the observational point of view, is the
mass corresponding to a given spherical overdensity, $\delta$, relative to
the mean density at the cluster redshift. Jenkins et al.\ considered two
values of $\delta$, 180 and 324 (the latter value corresponds to the virial
radii in the $\Lambda$CDM model with $\Om=0.3$). We measured baryon masses
for $\delta=324$. For this overdensity, Jenkins et al.\ obtain $A=0.316$,
$b=0.67$, and $c=3.82$.

If the baryon fraction in clusters is indeed universal, the baryon and total
overdensities must be equal, $\delta=\delta_b$, which allows one to use the
measured baryon mass $M_{b,324}$ as a proxy for the total mass, $M_{324}$.
Recall that the model for the baryon and total mass function are related via
equation (\ref{eq:Fb-Ft}).

Using the machinery described above, we can compute the model for the baryon
mass function for any set of cosmological parameters\footnote{We used the
  code kindly provided by A.~Jenkins. The code was slightly modified to
  include the power spectrum model by Eisenstein \& Hu (1998).} and then fit
it to the observations, self-consistently rescaling the observed mass
function to the cosmological parameters being tried. In the rest of this
section, we consider separately the effect of each cosmological parameter on
the $F_b(M_b)$ model and on scaling of the observed mass function. We also
consider the effect of possible deviations of $M_b/M_{\text{tot}}$ from the
universal value.

\subsection{Parameter \Om}

Parameter \Om{}, mainly through the product $\Omh$, determines the shape of
the transfer function $T(k)$ and therefore the slope of the mass
function. This is a strong effect which allows us to derive $\Om$ from the
cluster mass function.

In addition, \Om{} affects the growth of density perturbations. This effect
should be taken into account, because the observational mass function is
derived in a redshift interval of finite width. Therefore, we computed the
mass function models at several redshifts and weighted them with the survey
volume for the given mass.

The effect of $\Om$ on the observed mass function via the mass and volume
calculations is very weak and can be ignored.  The growth factor also
depends on $\Lambda$, but very weakly at low redshift, and therefore we
fixed $\Lambda=0.7$.

\subsection{Power Spectrum Normalization $\sigma_8$}
 
The normalization of the mass function model is exponentially sensitive to
$\sigma_8$, which allows a precise determination of this parameter from the
observed cluster abundance. 
 
The baryon mass measurements do not depend on $\sigma_8$, so there is no
associated scaling of the observed mass function.

\subsection{Hubble Constant $h$}

The derived baryon masses are very sensitive to the value of $h$,
$M_{b,324}\propto h^{-2.25}$ (eq.~\ref{eq:m:h}), and so the observed
mass function must be scaled accordingly. In the mass function model, the
Hubble constant enters through the product $\Omh$ (see above) and also
changes the mass scale, $M\propto h^{-1}$ (see e.g.\ Jenkins et al.). The
normalization of the model mass function scales as $h^3$, and, obviously,
the observed mass function scales identically.

Although the Hubble constant measurements now converge to $h=0.71$ (Freedman
et al.\ 2001, Spergel et al.\ 2003), our method is quite sensitive to
$h$. To determine the appropriate scalings, we varied $h$ in the range from
0.5 to 0.8. We use $h=0.71$ for a baseline model.

\subsection{Average Baryon Density $\Omega_b h^2$}

The average baryon density in the Universe is known quite accurately,
$\Omega_b h^2 = 0.0224\pm0.0009$ (Spergel et al.\ 2003, see also Burles et
al. 2001). This parameter enters the measurement of the baryon masses
$M_{b,324}$ (\S\,\ref{sec:mgas}, eq.~(\ref{eq:m:rhob})), the scaling between
the models for the total and baryon mass functions (eq.~\ref{eq:Fb-Ft}),
and, weakly, the shape of the transfer function. Variations of $\Omega_b
h^2$ within the above uncertainty range result in negligible changes in
either the observed or theoretical mass function. We, however, allowed
$\Omega_b h^2$ to vary in a broader range, from 0.018 to 0.025. If not
stated explicitly, all the results below are reported for $\Omega_b
h^2=0.0224$.

\subsection{Primordial Power Spectrum Slope $n$}

The observed mass function is independent of the power spectrum. For the
model mass function, the effect of $n$ is indistinguishable from that of the
product $\Omh$ because our entire mass range corresponds to a narrow (factor
of 2) wavenumber range, where the power spectrum of density perturbations is
indistinguishable from a power law. Our parameter constraints below were
generally obtained for $n=1$, expected in the inflationary models
(Starobinskii 1982, Guth \& Pi 1982, Hawking 1982) and favored by the recent
CMB observations (Spergel et al.\ 2003). The $n$-dependence of the derived
parameter values was studied by varying $n$ in the range from 0.85 to 1.15.

\subsection{Non-Universality of the Baryon Fraction $\Upsilon$}

Our method depends on the assumption that the baryon fraction in clusters is
universal, $M_b/M_{\text{tot}} = \Omega_b/\Om$. Any large deviations from
the universality may have serious observational and methodological
implications, and therefore require special attention.

Consider first how the baryon mass measurements have to be scaled, if the
baryon fraction is non-universal. Let us assume that at large radii, baryons
still follow the dark matter, i.e.\ $M_b/M_{\text{tot}} = \Upsilon\,
\Omega_b/\Om$ where $\Upsilon$ is a constant for each cluster. The first
obvious correction is to multiply the baryon mass by
$\Upsilon^{-1}$. However, if $\Upsilon\ne1$, the baryon and total
overdensities at each radius are different, $\delta_b=\Upsilon \delta$. We
need the baryon mass which corresponds to the \emph{total} overdensity
$\delta=324$, and therefore the baryon mass measurements must be corrected
by an additional factor of $\Upsilon^{-0.5}$ (cf.\
eq.~\ref{eq:m:delta}). Overall, the baryon mass measurement must be scaled
by a factor of $\Upsilon^{-1.5}$.

Consider now the effect of possible variants for the non-universality of the
cluster baryon fraction. Most of the deviations of $\Upsilon$ from unity can
be factorized into the following three possibilities:

1) $\Upa \ne 1$, but is the same for all clusters. Such a deviation accounts
for any systematic under- or overabundance of baryons in clusters. Such a
deviation is often observed in numerical experiments (Mathiesen et al.\ 1999
and references therein). Any systematic bias in the X-ray based gas mass
measurement is also indistinguishable from $\Upsilon\ne1$. For example,
Mathiesen et al.\ show that a systematic 5--10\% overestimation of the gas mass
is possible because of deviations from spherical symmetry. Overall, 15\%
variations of $\Upsilon$ seem to include the range for all known effects. We
studied the dependence of our results on such a systematic change of
$\Upsilon$ by repeating all the analyses for $\Upa = 0.85$, 1.0, and 1.15.

2) $\Upsilon=1$ on average, but there is some cluster-to-cluster scatter.
This can be represented by a convolution of the baryon mass function model
with the appropriate kernel. We find, that the effect on the derived
parameters is negligible as long as the scatter of $\Upsilon$ remains small
($\lesssim 15\%$), which is supported observationally (Mohr et al.\
1999).

3) \Upa\ is a function of the cluster mass and $\Upsilon\rightarrow 1$ in
massive clusters. This type of variation of the baryon fraction is expected,
e.g., in the models which involve a significant preheating of the
intergalactic medium. As a toy model for $\Upsilon(M)$, we considered the
parameterization
\begin{equation}
  \label{eq:depl.law}
  \Upa(M_b) = 1 - \frac{M_*}{M_b},\,\, M_* = 0.7\times10^{13}\Msun,
\end{equation}
which describes the results of numerical simulations by Bialek et al.\
(2001) where the level of preheating was adjusted so that the simulated
clusters reproduce the observed $L_x-T$ relation. This relation also is
reasonably consistent with recent direct measurements of the baryon fraction
in a small number of clusters spanning a temperature range of 1.5--10~keV
(Pratt \& Arnaud 2002, 2003; Sun et al.\ 2003, Allen et al.\ 2002).

Equation~[\ref{eq:depl.law}] was used as our baseline model in deriving the
parameter constraints, but we also considered all of the above possibilities
for deviations of $\Upsilon$ from unity.

The results of this section can be briefly summarized as follows. Given a
set of parameters $\Om$, $h$, $\sigma_8$, $n$, $\Omega_b$, we compute the
linear power spectrum of density perturbations (Eisenstein \& Hu 1998), then
convert it to the model for the total mass function (Jenkins et al. 2001)
and apply to the observed baryon mass function scaled along the mass axis as
follows (here, $\omega_b = \Omega_b h^2$)

\begin{equation}\label{eq:mb:mtot:bigscaling}
M h^{-1} = \Upa^{-1.5} M_b
\frac{\Om h^2}{\omega_b} \biggl(\frac{h}{0.71}\biggl)^{-2.25}
\biggl(\frac{\omega_b}{0.0224}\biggl)^{-0.5}
\end{equation}

\section{Results}
\label{sec:results}

\subsection{Application of the Models to the Data}

In the previous section, we discussed how to compute a model for the baryon
mass function for a given set of CDM model parameters. Parameter constraints
can be obtained by fitting such models to the observed mass function using
the Maximum Likelihood (ML) method. Note, however, that an exact
computation of the likelihood in our case is rather cumbersome because of
the presence of the mass measurement uncertainties in addition to the
Poisson noise (\S\,\ref{sec:mfun}). To avoid the computational overhead, we
used a simpler technique to derive the cosmological parameter
constraints. We use 68\% and 95\% confidence bands for the mass function
obtained in \S\,\ref{sec:mfun} using an ML fit to the Schechter function. A
mass function model which is entirely within a 68\% confidence band in the
interesting mass range is deemed acceptable at the 68\% confidence level
(and likewise for the 95\% confidence level). Further comparison of this and
ML methods is given in Vikhlinin et al.\ (2003).

This method results in slightly more conservative confidence intervals than
a direct ML fitting. An advantage of our approach over ML fits is its direct
relation to the goodness-of-fit information. By design, all acceptable mass
function models actually fit the data. On the contrary, the goodness-of-fit
information is generally unavailable with the ML approach and sometimes it
is possible to obtain artificially narrow confidence intervals on parameters
only because the best fit model provides a poor fit to the data.

For parameter constraints, we considered the observed baryon mass function
in a restricted mass range, $4\times10^{13}\Msun\le M_b\le
3\times10^{14}\Msun$. The choice of the lower limit is motivated by the
possibility that the baryon fraction drops systematically in the less
massive clusters (see above). Also, the number of clusters in our sample
with $M_b<4\times10^{13}\Msun$ is small and so the effects of incompleteness
may be important. The upper limit of the mass interval was chosen to exclude
the two most massive and potentially peculiar clusters A2142 and A2163 (Fig.~\ref{fig:cumfunc}).

\begin{figure}
  \vspace*{-5mm}
  \centerline{
    \includegraphics[width=0.99\linewidth]{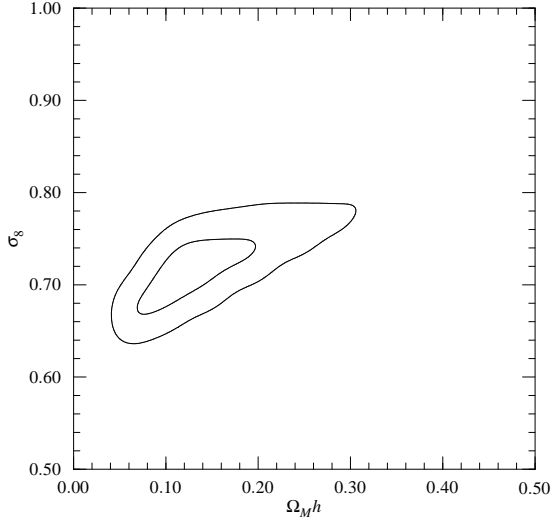} } \vspace*{-7mm}
  \caption{Confidence intervals (68\% and 95\%) for the baseline model
    ($h=0.71$, $n=1$, $\Omega_b h^2=0.0224$, and $\Upsilon$ from
    eq.~[\ref{eq:depl.law}]).}
  \label{fig:1-2sig.contour}
\end{figure}

As was discussed in \S\,\ref{sec:theory}, the most sensitive constraints
from the cluster baryon mass function are for parameters $\sigma_8$ and
$\Omh$. Other parameters, $h$, $n$, $\Omega_b h^2$, cannot be constrained
from such an analysis. Their variation causes simple scalings of the
constraints on $\sigma_8$ and $\Omh$. Below, we will describe constraints on
$\sigma_8$ and $\Omh$ for the baseline model, $h=0.71$, $n=1$, $\Omega_b
h^2=0.0224$, and provide scalings of these constraints corresponding to
variations of the latter three parameters as well as deviations of
$\Upsilon(M)$ from the adopted model. We consider separately the
$\sigma_8-\Omh$ constraints from fitting the shape of the mass function in
the $4\times10^{13}\Msun\le M_b\le 3\times10^{14}\Msun$ mass range and from
using only the normalization of the mass function at $M_b=1.0\times10^{14}\,M_\odot$.

\subsection{Fitting the Shape of the Baryon Mass Function}

A simultaneous determination of the parameters $\sigma_8$ and $\Omh$ is
possible from fitting the mass function in a broad $M$-range because the
slope of the mass function is sensitive to $\Omh$ and its normalization ---
to $\sigma_8$. In Fig.~\ref{fig:1-2sig.contour}, we show the 68\% and 95\%
confidence regions for our baseline model. The best fit is obtained for $\Om
h=0.13$ and $\sigma_8=0.72$ (the corresponding mass function model is shown
by the solid line in Fig.~\ref{fig:cumfunc}). The 68\% uncertainty interval
on $\Om h$ includes the `Cosmic Concordance' value of $\Om=0.3$, but the
confidence regions are skewed towards lower values of $\Om$. This result is
similar to a conclusion from the cluster temperature function studies that
the cluster mass function is shallower than predicted in CDM (Henry 2000).

\emph{Parameter $h$}: Consider now how the $\sigma_8-\Omh$ constraints
depend on the exact value of $h$. The effect of the Hubble constant on the
shape of the power spectrum is important only through the product
$\Omh$. Therefore, the effect of $h$ is to shift the confidence levels in
Fig.~\ref{fig:1-2sig.contour} along the $\sigma_8$ axis. Experimenting with
different values of $h$, we find that this shift is
\begin{equation}
  \label{eq:s8(h)}
  \Delta\sigma_8  = 0.31 \Delta h.
\end{equation}

Note that the $\sigma_8$ constraints derived from the cluster temperature
function are independent of $h$ because both the mass corresponding to a
given $T$ and the independent variable of the model mass function scale as
$h^{-1}$. In our case, a different scaling of the baryon mass measurement
with $h$ ($M_b\propto h^{-2.25}$, see \S\,\ref{sec:theory}) leads to a
dependence of the derived $\sigma_8$ on $h$.

\emph{Parameter $n$}: The shape of the cluster mass function is defined by
the slope of the linear power spectrum of density perturbations in the
narrow wavenumber range around $k\sim0.1h$, where $P(k)$ is accurately
approximated by a power law. This means that we cannot disentangle the slope
of the primordial (inflationary) power spectrum $n$ from the slope of the
transfer function on cluster scales. For example, for a higher value of $n$
we need a flatter transfer function, i.e., a lower value of
$\Omh$. Therefore, our $\sigma_8-\Omh$ constraints computed for $n=1$
(Fig.~\ref{fig:1-2sig.contour}) simply move along the $\Omh$ axis when the
assumed value of $n$ is changed:
\begin{equation}
\Delta (\Omh) = -0.21\Delta n.
\end{equation}

\begin{figure}
  \vspace*{-5mm}
  \centerline{\includegraphics[width=0.99\linewidth]{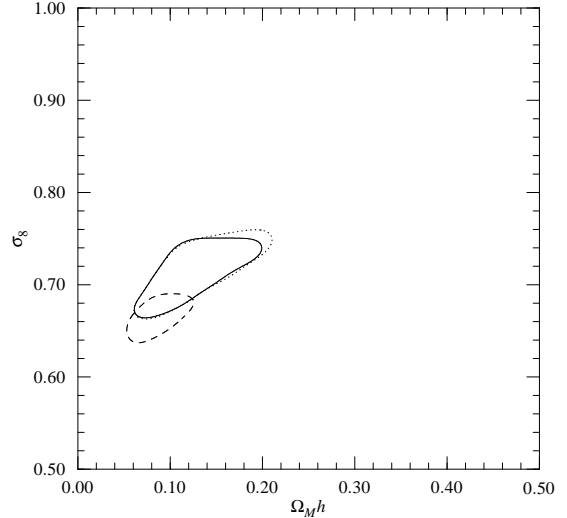}}
  \vspace*{-7mm}
  \caption{The effect of variations in the baryon fraction on the cosmological
    constraints. The solid contour corresponds to \Upa\ given by eq.\
    [\ref{eq:depl.law}] (baseline model), the dotted contour is obtained for
    the case of a $10\%$ cluster-to-cluster scatter in \Upa\ around the mean
    relation [\ref{eq:depl.law}], and the dashed contour corresponds to
    $\Upa = 1$.}
  \label{fig:dublcont}
\end{figure}

\emph{Parameter $\Omega_b h^2$}: This parameter affects the mass function model
via the transfer function, but the effect is miniscule for reasonable values
of $\Omega_b/\Om$. The effect of the average baryon density on the
observed baryon mass function (via the scaling of $M_b$,
eq.~\ref{eq:m:rhob}) is much stronger. This effect is equivalent to the
$h$-scaling of $M_b$ considered above --- it shifts the $\sigma_8-\Omh$
confidence intervals along the $\sigma_8$ axis,
\begin{equation}
\Delta\sigma_8 = -18.0\, \Delta\Omega_b h^2
\end{equation}

\emph{Non-universal baryon fraction, \Upa}: As we noted before, any large
deviations of the cluster baryon fraction from the universal value
$\Omega_b/\Om$ can significantly affect our results.

The simplest case is a uniform scaling of the baryon fraction by a factor
$\Upsilon_0$, i.e.\ $\Upsilon(M) = \Upsilon_0 (1-M_*/M_b)$ for our baseline
model (eq.~\ref{eq:depl.law}) or simply $\Upsilon(M) = \Upsilon_0$. Such a
variation in $\Upsilon$ simply shifts the observed baryon mass fraction
along the mass axis by a factor of $\Upsilon_0^{1.5}$
(eq.~\ref{eq:mb:mtot:bigscaling}). The effect on the $\sigma_8-\Omh$
confidence intervals is the same as in the case of variations of $\Omega_b
h^2$ and $h$ --- the confidence intervals shift along the $\sigma_8$ axis,
\begin{equation}
  \Delta \sigma_8 = -0.37\Delta \Upa_0,
\end{equation}
and the constraints on $\Omh$ are unaffected.

If $\Upsilon$ is a function of the cluster mass, as in (\ref{eq:depl.law}),
the shapes of the total and baryon mass functions are different (cf.\
eq.~\ref{eq:mb:mtot:bigscaling}). Therefore, the derived values of both
$\sigma_8$ and $\Omh$ are affected. The results of fitting the observed mass
function assuming either $\Upsilon(M)$ given by equation (\ref{eq:depl.law})
or $\Upsilon(M)=1$ are shown in Fig.~\ref{fig:dublcont}. For
$\Upsilon(M)=1$, we obtain lower values of both $\Omh$ and $\sigma_8$
(dashed contour), although the effect on $\sigma_8$ is partly due to the
correlation of this parameter with the assumed value of $\Omega_M$ (see
below). If the baryon fraction decreases in low-mass clusters, the
cumulative baryon mass function is progressively shifted towards smaller
$M$. This makes the predicted $F(M_b)$ flatter with a smaller normalization
than for $\Upsilon=1$. Therefore, the data will require steeper models for
the total mass functions (i.e., higher values of $\Omh$) with a higher
normalization (i.e., higher values of $\sigma_8$).

Dotted contour in Fig.~7 shows the effect of a 10\% cluster-to-cluster
scatter in $\Upsilon$ on the $\sigma_8-\Omh$ constraints. As expected, the
scatter has a small effect on the results, only to expand the size of the
confidence interval by $\sim 12\%$.

\begin{figure}
  \vspace*{-5mm}
  \centerline{ \includegraphics[width=0.99\linewidth]{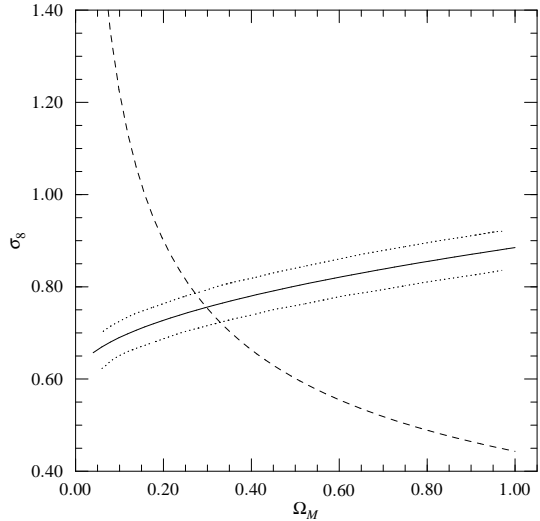} }
  \vspace*{-7mm}
  \caption{
    Constraints on $\sigma_8$ using only the normalization of the baryon
    mass fraction and assuming $h=0.71$. The dotted lines correspond to the
    68\% confidence region and the solid line shows the analytic
    approximation (eq.~[\ref{eq:sigma8:vs:Omega}]). The dashed line shows a
    typical $\sigma_8(\Omega_M)$ dependence obtained from the normalization
    of the cluster temperature function (adapted from Seljak 2002).}
  \label{fig:lane}
\end{figure}

\subsection{Constraints from Normalization of the Baryon Mass Function}

The $\sigma_8$ constraints are often derived from the cluster abundance data
by using only the normalization of the mass or temperature functions, e.g.,
the number density of $T>6$~keV clusters.  We have applied this approach to
our baryon mass function data. The number density of clusters with
$M_b>10^{14}\Msun$, the median mass in the sample, was used as the only
observable. This mass threshold approximately corresponds to clusters with
$T=5.5$~keV (from the $M_{b,324}-T$ relation presented in Voevodkin et al.\
2002). Since no information about the shape of the mass function is used, it
is possible to constrain only $\sigma_8$ as a function of $\Om$. The results
are shown in Fig.~\ref{fig:lane} by dotted lines. The allowed band (68\%
confidence) of $\sigma_8$ for our baseline model is accurately described by
the following equation (solid line in Fig.~\ref{fig:lane}):
\begin{equation}\label{eq:sigma8:vs:Omega}
  \sigma_8 (\Om) = 0.60+ 0.28\,\Om^{0.5}\pm0.04.
\end{equation}

The $\sigma_8$ constraints from the normalization of the cluster temperature
or total mass functions are also $\Om$-dependent. Typically, one finds
$\sigma_8\propto \Om^{-\alpha}$ with $\alpha=0.4-0.6$ from these
analyses. Our $\Om$-dependence is much weaker (solid vs.\ dashed line in
Fig.~\ref{fig:lane}). The dependence of $\sigma_8$ on $\Om$ in the baryon
mass function analysis is weak, because the relation between the cluster
baryon mass and the corresponding linear scale is $\Om$-independent,
$\lambda^3 \sim M_b/\langle\rho_b\rangle$, while for the total mass,
$\lambda^3 \sim M_{\text{tot}}/(\Om\,\rho_{\text{cr}})$. This is one of the
advantages of our method. The price we have to pay is the introduction of
new dependencies for the derived value of $\sigma_8$ on parameters $h$,
$\Omega_b\,h^2$, and $\Upsilon$. Table~\ref{tab:scale.rel} summarizes the
changes in the coefficients of equation~(\ref{eq:sigma8:vs:Omega}) as these
parameters are varied.

\begin{table}[t]
  \def\tnote#1{\ensuremath{^{\text{#1}}}}
  \def\d{\phantom{1}}
  \def\arraystretch{1.1}
  \centering
  \caption{Variations of $a$ and $b$ in the relation $\sigma_8(\Om) =
    a+b\Om^{0.5}$ as the main parameters are changed.}\label{tab:scale.rel}
    \def\arraystretch{1.15} \footnotesize
  \begin{tabular}{ccc}
    \hline
    \hline
    \multicolumn{1}{c}{Parameter} & $\Delta a$ & $\Delta b$\\
    \hline

    $h$                          & $0.15$  & $0.47$ \\
    $\Omega_b h^2$               & $-6.17$ & $-23.7$\\
    $\Upa$ from eq.~(\ref{eq:depl.law}) & $-0.08$ & $-0.63$\\
    $\Upa=\mathrm{const}$               & $-0.12$ & $-0.51$ \\

    \hline
  \end{tabular}
\end{table}

\section{Comparison with earlier work}\label{sec:comparison}

Determination of $\sigma_8$ from various datasets at $z=0$ is a
well-established field and so a comparison of our results with at least some
recent relevant studies is in order.

Our results can be most directly compared with the determination of
$\sigma_8$ from the cluster abundance data. Most such studies use the
cluster temperature as a proxy for the total mass and a normalization of the
local temperature function at $T=5-7$~keV as the primary observable. A
review of recent results from the temperature function can be found in
Pierpaoli et al.\ (2003). As we noted before, the scatter in the derived
values of $\sigma_8$ in this method is mostly due to different assumptions
regarding the normalization of the $M-T$ relation. Generally, our
determination of $\sigma_8$ is in excellent agreement (for $\Omega=0.3$)
with those studies that use the $M-T$ normalization from the X-ray total
mass measurements assuming hydrostatic equilibrium for the ICM (e.g.,
Markevitch 1998, Seljak 2002). Our $\sigma_8$ is lower than the
determination from the optical cluster mass function by Girardi et al.\
(1998). However, our value (for $\Omega_M=0.3$) is consistent with the
analyses of Bahcall et al.\ (2003) who used the cluster mass function
estimated from the Sloan Digital Sky Survey, and of Melchiorri et al.\
(2003) who combined the SDSS results with the pre-\emph{WMAP} CMB
measurements.

Two papers have used essentially the local X-ray luminosity function (XLF)
and the X-ray normalized $L_x-M_{\text{tot}}$ relation (Reiprich \&
B\"ohringer 2002, Allen et al.\ 2003) to estimate the total mass
function. Our results on $\sigma_8$ are in a good agreement with these works
(for $\Omega=0.3$).  Reiprich \& B\"ohringer also fit the shape of their
mass function to derive $\Omh=0.09\pm0.03$. This is consistent but on the
low side of our confidence intervals (Fig.~\ref{fig:1-2sig.contour}), likely
because Reiprich \& B\"ohringer used the Press-Schechter approximation for
the mass function in their modeling which is less accurate than the Jenkins
et al.\ fit employed here.

Matter density fluctuations can be traced by the spatial distribution of
galaxies. Two modern surveys, SDSS and 2dF (York et al.\ 2000, Colless et
al.\ 2001) provide excellent datasets for these studies. The galaxy power
spectrum from 2dF implies $\Omh=0.20\pm0.03$ (Percival et al.\ 2001),
consistent with our results for the baseline model.

Using clusters as tracers of large scale structure is also a promising
approach to determine the matter power spectrum. Using the power spectrum of
426 clusters from the REFLEX survey (B\"ohringer et al.\ 2001), Schuecker et
al.\ (2003) determine $\sigma_8=0.71\pm0.03$ and $\Omh=0.24\pm0.02$. Note
that this method is probably less sensitive to the normalization of the
$M_{\rm tot}-L_x$ relation than using the normalization of the XTF or XLF
because the bias factor changes more slowly than the spatial density of
clusters as a function of mass (Mo \& White 1996). It is encouraging that
our results are in good agreement with the Schuecker et al.\ values.

Finally, $\sigma_8$ can be determined from the fluctuations of the cosmic
microwave background. The recent \emph{WMAP} measurements of the CMB angular
fluctuations and Thompson optical depth of the Universe indicate
$\sigma_8=0.9\pm0.1$ assuming pure CDM, $w=-1$, and some other common
cosmological priors (Spergel et al.\ 2003); this is higher than, but
consistent with, our value.

\section{Summary and Conclusions}

We have derived the baryon mass function for galaxy clusters at a median
redshift $\langle z\rangle=0.05$. The baryon (gas+stars) mass is measured
within a radius of mean baryon overdensity $\delta=324$, using direct
deprojection of the ROSAT X-ray imaging data (for gas mass) and an
established correlation between $M_{\rm g}$ and $L_{\rm opt}$ (for stellar
mass).

We argue that if the baryon fraction in cluster is nearly universal, the
baryon mass function is an excellent proxy for the total mass function ---
$F(M_{\rm tot})$ is obtained from $F_b(M_b)$ simply by a constant log shift
along the mass axis. In fact, $F_b(M_b)$ contains enough information to
derive the amplitude and slope of the density fluctuations at cluster
scales, even if the absolute value of $M_{\rm tot}/M_b$ is unknown.

Using this method we obtain $\sigma_8=0.72\pm0.04$ and
$\Omh=0.13\pm0.07$. The baryon mass function method has an important
advantage in that it does not rely on rather uncertain observational
determinations of the total cluster mass. The tradeoff is the sensitivity of
the derived values of $\sigma_8$ and $\Omh$ to other cosmological parameters
(such as $h$, $\Omega_b$). Our method is also sensitive to any significant
deviations of the cluster baryon fraction from universality
(\S\,\ref{sec:results}). However, the good agreement of our results with a
large number of measurements by other, independent methods
(\S\,\ref{sec:comparison}) shows that our assumptions about the weak trends
in the baryon fraction are reasonable. We hope that future progress in
observations and physical models of the ICM will reduce this uncertainty to
such a level that easily measurable baryon masses can be used as a reliable
proxy for the total mass at all redshifts.

\acknowledgements

We thank A.~Jenkins for providing the software for computing the model mass
functions, and M.~Markevitch and C.~Jones for careful reading of the
manuscript.  This work was supported by Russian Basic Reseach Foundation
(grants 02-02-06658 and 00-02-17124) and by NASA grant NAG59217 and contract
NAS8-39073. A.~Voevodkin thanks SAO for hospitality during the course of
this research and also aknowledges support from the Russian Academy of
Sciences via the Young Scientist program.

\end{document}